\documentclass[11pt]{article}

\usepackage{amsmath,amssymb,amsfonts}
\usepackage{amsthm}
\usepackage{graphicx}
\usepackage{booktabs}
\usepackage{multirow}
\usepackage{natbib}
\usepackage{geometry}
\usepackage{setspace}
\usepackage{enumitem}
\usepackage{hyperref}
\usepackage{bm}
\usepackage{epsfig}
\usepackage{cite}
\usepackage{algorithm, algorithmic}  
\usepackage{mathtools}  

\usepackage{enumerate}
\usepackage{url} 
\usepackage{xcolor}

\newtheorem{theorem}{Theorem}
\newtheorem{corollary}{Corollary}

\newtheorem{proposition}{Proposition}
\newtheorem{lemma}{Lemma}
\theoremstyle{definition}
\newtheorem{definition}{Definition}

\newtheorem{example}{Example}
\newtheorem{remark}{Remark}

\newcommand{\beq}{\begin{equation}}
\newcommand{\eeq}{\end{equation}}
\newcommand{\beas}{\begin{eqnarray*}}
\newcommand{\eeas}{\end{eqnarray*}}
\newcommand{\bea}{\begin{eqnarray}}
\newcommand{\eea}{\end{eqnarray}}
\newcommand{\bei}{\begin{itemize}}
\newcommand{\eei}{\end{itemize}}
\newcommand{\ben}{\begin{enumerate}}
\newcommand{\een}{\end{enumerate}}
\newcommand{\bet}{\begin{theorem}}
\newcommand{\eet}{\end{theorem}}
\newcommand{\bel}{\begin{lemma}}
\newcommand{\eel}{\end{lemma}}
\newcommand{\bep}{\begin{proposition}}
\newcommand{\eep}{\end{proposition}}
\newcommand{\bed}{\begin{definition}}
\newcommand{\eed}{\end{definition}}
\newcommand{\bec}{\begin{corollary}}
\newcommand{\eec}{\end{corollary}}
\newcommand{\bex}{\begin{example}}
\newcommand{\eex}{\end{example}}

\newcommand{\argmin}{\mathop{\rm arg\min}}

\def\X{{\bm x}}
\def\Z{{\bm z}}
\def\z{{\bm z}}
\def\B{{\bf B}}

\def\bbeta{{\boldsymbol \beta}}
\def\bpi{{\boldsymbol \pi}}
\def\bPi{{\boldsymbol \Pi}}

\title{\bf Targeted learning via probabilistic subpopulation matching}
\author{Xiaokang Liu\thanks{Co-first author. Department of Statistics and Data Science, University of Missouri, xiaokang.liu@missouri.edu} \and  Jie Hu \thanks{Co-first author. Department of Biostatistics, Epidemiology and Informatics,
    University of Pennsylvania, jie.hu@pennmedicine.upenn.edu} \and Naimin Jing\thanks{Biostatistics and Research Decision Sciences, Merck \& Co., Inc, naiminjing@gmail.com} \and Yang Ning\thanks{Department of Statistics and Data Science,
       Cornell University, yn265@cornell.edu}
       \and  Cheng Yong Tang\thanks{Department of Statistics, Operations, and Data Science, 
    Temple University, yongtang@temple.edu} \and Runze Li\thanks{Department of Statistics, 
    The Pennsylvania State University, ril4@psu.edu}
       \and Yong Chen\thanks{Corresponding author. Department of Biostatistics, Epidemiology and Informatics,
       University of Pennsylvania, ychen123@upenn.edu}}
\date{}

\begin{document}
\maketitle

\begin{abstract}
In biomedical research, to obtain more accurate prediction results from a target study, leveraging information from multiple similar source studies is proved to be useful. However, in many biomedical applications based on real-world data, populations under consideration in different studies, e.g., clinical sites, can be heterogeneous, leading to challenges in properly borrowing information towards the target study. The state of art methods are typically based on study-level matching to identify source studies that are similar to the target study, whilst samples from source studies that significantly differ from the target study will all be dropped {\em at the study level}, which can lead to substantial loss of information. We consider a general situation where all studies are sampled from a super-population composed of distinct subpopulations, and propose a novel framework of targeted learning via subpopulation matching. In contrast to the existing study-level matching methods, measuring similarities between subpopulations can effectively decompose both within- and between-study heterogeneity, allowing incorporation of  information from all source studies without dropping any samples as in the existing methods. We devise the proposed framework as a two-step procedure, where a finite mixture model is first fitted jointly across all studies to provide subject-wise probabilistic subpopulation information, followed by a step of within-subpopulation information transferring from source studies to the target study for each identified subpopulation. We establish the non-asymptotic properties of our estimator and demonstrate the ability of our method to improve prediction at the target study via simulation studies. 
\end{abstract}


\newpage

\section{Introduction}
\label{sec:intro}

The World Health Organization officially declared the end of COVID-19 as a global public health emergency on May 5, 2023. However, the Post-Acute Sequelae of SARS-CoV-2 infection (PASC), commonly known as long COVID, remains a global threat to the public due to its significant impact on long-term health outcomes and healthcare resource planning. PASC includes a wide range of symptoms that persist or newly emerge after the acute phase of COVID-19 infection, lasting for weeks or months and affecting multiple body systems \citep{davis2023long, jama-pasc}. Predicting PASC is critical yet challenging, as its development can be influenced by various factors, including the severity of acute phase infection, underlying health conditions, and demographic factors \citep{ pfaff2022identifying}. 
Incorporating all these factors into prediction would result in a complex and high-dimensional model, which typically requires a large sample size for reliable prediction. In an electronic health records (EHRs)-based study aimed at predicting the risk of developing PASC in children with SARS-CoV-2 infection at a children’s hospital,  
the available sample size is limited 
after applying strict eligibility criteria. These criteria included a confirmed COVID-19 infection and at least two years of medical history, ensuring the clinical relevance and reproducibility of the findings.
Suffering from the limited sample size of a well-defined clinical cohort, the prediction accuracy could be suboptimal at the target hospital. A promising solution lies in leveraging information from 
other peer collaborating 
hospitals (hereafter, referred to as source studies) with relevant PASC-related patient data, to enhance the prediction models built at the target study.

When 
all source studies share the same prediction model with the target study (i.e., there is no between-study heterogeneity), pooling data from all source studies and the target study is a viable solution to address the issue of limited sample size. 
In the above PASC study, 
{several collaborating hospitals are available in addition to the target study, enabling a substantial expansion of the sample size} 
if we pool all the patient-level data together. 
However, in practice, various factors, encompassing organizational, operational, and environmental aspects, are contributing to the differences between hospitals, leading to non-negligible between-study heterogeneity. For example, the location of a hospital may affect the distribution of patient demographics, and the size, resources, and staffing of a hospital could impact its quality of care as well. Ignoring the differences between the source studies and the target study may make the resulting prediction model severely deviate from the target model. Therefore, there is an urgent need to develop methods for understanding how, and to what extent, we can learn from source studies in the presence of between-study heterogeneity.


Transfer learning is a technique that improves the predictive performance of a statistical model on a target study by leveraging information, patterns, or knowledge derived from source studies \citep{pan2009survey}. Over the past few years, statistical transfer learning has seen rapid development across various learning tasks, where theoretical properties have been established. For example, \citet{li2020transfer} proposed algorithms for enhancing a high-dimensional linear regression on the target study by transferring information from related auxiliary studies and established minimax optimality. \citet{li2020transferggm} proposed a high-dimensional Gaussian graphical model and proved a faster convergence rate than the minimax rate derived from the target study alone. \citet{cai2021transfer} considered minimax transfer learning for nonparametric classification and developed a data-driven adaptive classifier. 
\citet{tian2022transfer} independently proposed an innovative transfer learning approach for high-dimensional generalized linear models, highlighting its efficiency. In the federated setting where patient-level data sharing is prohibited, \citet{li2021targetingftl} introduced a novel federated transfer learning approach for high-dimensional regression
models, and \citet{li2023accommodating} designed an innovative Cox model transfer learning approach where the source study only needs to share summary statistics with the target study. 

A key step in transfer learning is assessing the similarity between a source study and the target study to ensure that knowledge in a source study is transferable to the target study.
This strategy can effectively avoid the negative transfer effects \citep{li2020transfer, tian2022transfer}, and is well-suited for borrowing information across genetic/genomic studies, when there is a large number of source studies with small to moderate sample sizes but some of them are with high between-study heterogeneity. In such cases, dropping less relevant studies is protective to the target study.  
However, in the context of distributed research networks that involve different hospitals, where the number of source studies (i.e., hospitals) is often limited and each usually contains hundreds of thousands of patients, excluding any hospital could result in substantial information loss. 
It is important to propose a new framework in this new setting to retain all source studies in the analysis while safeguarding the prediction model against negative transfers.

In this paper, we propose one of the strategies, which is to decompose the between-study heterogeneity by a subpopulation structure. Subtyping, also known as subphenotyping, is a clustering procedure to find distinct subpopulations from a complex population, and is commonly used in biomedical studies to facilitate better understanding of disease development mechanisms \citep{zhang2023data}. In PASC study, subtyping can be used to categorize all patients into several subpopulations with different baseline health status, which may affect how risk factors influence the development of PASC. For example, patients with pre-existing conditions such as diabetes and cardiovascular diseases might be more susceptible to PASC compared to healthier patients, leading to significantly different risk factor profiles across multiple health statuses. 
Additionally, due to variations in factors such as location and size between hospitals, the mixing proportions of subpopulations at one hospital can differ significantly from those at other hospitals. 
These subpopulation-specific disease development mechanisms, along with study-specific subpopulation mixing proportions, can largely explain the observed non-negligible between-study heterogeneity.

In this article, inspired by the idea of heterogeneity decomposition, we devise our method as a two-step procedure to integrate subpopulation structure into the analysis. 
First, we apply a joint clustering analysis with data from all studies to identify the shared subpopulations, and use study-specific subpopulation mixing proportions 
to take account the majority of between-study heterogeneity. Then, information transfer is conducted within these subpopulations, 
enabling the utilization of information from all source studies. Throughout the process, each patient participates in the analysis through their subpopulation membership probabilities, which mitigates the impact of misclassification and 
results in what we term ``probabilistic subpopulation matching learning.'' 
Our approach not only facilitates the transfer of knowledge from all source studies but also provides subpopulation-specific assessments of risk factor importance through penalization,  decomposing both within- and between-study heterogeneity and offering more informative guidance for individuals with diverse profiles.

We provide a computational algorithm to implement all steps including subtyping, knowledge transfer, and prediction model training in a unified framework, and study the theoretical properties of the resultant estimator from our algorithm. 
Importantly, our theoretical analysis explored the impact of uncertainty in subpopulation structure estimation on the final regression parameter estimation. 
Our results indicated that if the estimation error of the subpopulation structure is well controlled, then the estimation error bound of our regression estimator matches the results in \citet{tian2022transfer} for the situation where all source studies are included in the informative set. 
Our simulation studies also demonstrated that the proposed method achieves higher estimation and prediction accuracy compared to traditional transfer learning methods and target analysis alone under various levels of between-study heterogeneity. 

The methodology is detailed in Section 2. Theoretical results are presented in Section 3. Simulation study results are presented in Section 4. 
Section 5 offers concluding remarks and discussions.

\section{Method}
\subsection{Notation}
We first introduce the notations to be used throughout the paper. We use bold capital letters (e.g., $\B$) to denote matrices and bold lowercase letters (e.g., $\bm{x}$) to denote vectors.
For a $p$-dimensional vector $\bm{x} = (x_1, \ldots, x_p)^T$, we write its $\ell_0$-norm as $\|\bm{x}\|_0=\#\left\{j: x_j \neq 0\right\}$, its $\ell_q$ norm as $\|\bm{x}\|_q = \left(\sum_{i=1}^p |x_i|^q\right)^{1/q}$ for $q \in (0,2]$, and its $\ell_{\infty}$ norm as $\|\bm{x}\|_{\infty} = \sup\limits_{j} |x_j|$. 
For a matrix $\B=(b_{ij})$, the $\ell_1$-norm is $\|\B\|_1=\sum_{i,j}|b_{ij}|$ and its $\ell_2$-norm is defined as $\|\B\|_2=(\sum_{i,j}b_{ij}^2)^{1/2}$. 
For a symmetric matrix $\bm{\Sigma}$, we denote its largest and smallest eigenvalues by $\lambda_{\max} (\bm{\Sigma})$ and $\lambda_{\min} (\bm{\Sigma})$, respectively.
For two nonzero real sequences $\left\{a_n\right\}_{n=1}^{\infty}$ and $\left\{b_n\right\}_{n=1}^{\infty}$, we use $a_n \ll b_n$ 
to represent $\left|a_n / b_n\right| \rightarrow 0$, and $a_n \lesssim b_n$ or $a_n = O(b_n)$ to represent $\sup_n |a_n / b_n| < \infty$. 
For two real numbers $a$ and $b$, $a \vee b$ and $a \wedge b$ represent $\max(a, b)$ and $\min(a, b)$, respectively. 
For two random variable sequences $\left\{x_n\right\}_{n=1}^{\infty}$ and $\left\{y_n\right\}_{n=1}^{\infty}$,  $x_n=O_p\left(y_n\right)$ means that for any $\epsilon>0$, there exists a positive constant $M$ such that $\sup _n P\left(\left|x_n / y_n\right|>M\right) \leq \epsilon$. 
For a function $g(x) \in \mathbb{R}$, we use $g^{\prime}(x)$ and $g^{\prime\prime}(x)$ to denote its first and second derivatives, respectively. Unless otherwise noted,  the expectation $\mathbb{E}$, variance $\mathrm{var}$, and covariance $\mathrm{cov}$ are calculated based on all sources of randomness. Furthermore, we use $c_0, c_1, c_2 \ldots$ to represent positive constants, which may vary depending on the context. 

\subsection{Setting}



For the rest of the paper, we will use the study of 
developing prediction models and identifying risk factors for PASC among COVID-19 infected children depending on patients' various pre-infection baseline health statuses as an example to provide a concrete context, explain our assumptions, and illustrate the idea.

In the target study, for the $i$-th patient with $i=1,\ldots,n_0$, we denote the scalar outcome, e.g., whether or not a patient developed PASC during the observation window, as $y_{0i}\in \mathbb{R}$. To facilitate the prediction of the outcome, from each patient we collect a set of potential risk factors $\X_{0i}=(x_{0i1},\ldots,x_{0ip})^T\in\mathbb{R}^p$ where $p$ can be large. 
Different baseline health status of patients may affect how risk factors influence the development of PASC, leading to significantly different 
relationships between $\X_{0i}$ and $y_{0i}$ across multiple health statuses. For example, patients with pre-existing conditions might be more susceptible to PASC compared to healthier patients, indicating the necessity of incorporating the diversity in pre-infection health status among patients into PASC analysis to provide more personalized insights for individuals dealing with the consequences of COVID-19 infection.
To infer the baseline health status for each patient, we collect an additional set of variables $\Z_{0i}=(z_{0i1},\ldots,z_{0iq})^T\in\mathbb{R}^q$,
and let $C_{0i}$ to be an unobserved indicator variable to denote the baseline health status.

Suppose there are in total $C$ subgroups of patients that present distinct baseline health status, where $C$ is assumed to be known. 
Then, given $C_{0i}=c$, $c=1,\ldots,C$, we consider a generalized linear model (GLM) with the canonical link to describe the effects from predictors on the outcome
\begin{align*}
f(y_{0i}|\bm{x}_{0i},C_{0i}=c;\bbeta_{0c}) = \exp\left\{ \frac{y_{0i}\bm{x}_{0i}^T\bbeta_{0c} - g(\X_{0i}^T\bbeta_{0c})}{a(\phi)} + c(y_{0i}, \phi) \right\},
\end{align*}
where the form of the functions $g(\cdot)$, $a(\cdot)$ and $c(\cdot)$ is determined by the outcome type, and the dispersion parameter $\phi$ is assumed to be known. 
The sample size of the target study $n_0$ can be small, and we aim to improve the estimation of $\B_0=(\bm{\beta}_{01}, \ldots, \bm{\beta}_{0C})\in\mathbb{R}^{p\times C}$ by transferring information from available source studies.

Suppose we have $K$ source studies. From each source study, we collect observations for the same set of variables as in the target study, i.e., in the $k$-th source study ($k=1,\ldots,K$), we have $\{(y_{ki}, \X_{ki}, \Z_{ki}, C_{ki})\}_{i=1}^{n_k}$, and the corresponding parameters are $\B_k=(\bm{\beta}_{k1}, \ldots, \bm{\beta}_{kC})\in\mathbb{R}^{p\times C}$. Due to the existence of between-study heterogeneity, we allow $\B_k \neq \B_j$ for any $k \neq j$ and $k, j \in \{0, 1, \ldots, K\}$ where $k=0$ denotes the target study. We use $\bm{\theta}_{reg}=(\B_0, \B_1, \ldots, \B_K)$ to denote all regression parameters involved in the analysis. 
Across all $K+1$ studies, multiple factors, including different locations, administration patterns, and referral patterns of the hospitals, are contributing to the between-study heterogeneity. If we measure similarity between the target study and source studies at the study level, 
some hospitals will be removed from the informative set from which the target study should borrow information. However, since the number of participating hospitals is relatively small and each hospital may contain observations from as many as tens of thousands of patients, removing any hospital from the analysis may lead to a substantial information loss.    

\subsection{A Probabilistic Subpopulation Matching Approach}

To address this challenge, our strategy  is to decompose the large between-study heterogeneity, thereby allowing for transferring information from all source studies. This procedure also decomposes the within-study heterogeneity for all studies. To illustrate our idea, consider the PASC analysis where the total population, or super-population, is consisting of children who had SARS-CoV-2 infection. For all the infected children, their baseline health status is believed to play a role in the development of PASC, which naturally divides the whole population into several subgroups. Within each subgroup, the patients share the same baseline health status $C_{ki}=c$ ($c=1,\ldots,C$) and also the way in which the risk factors influencing the development of PASC (i.e., $\bm{\beta}_{kc}$), while between the subgroups, the disease development mechanisms can be significantly different, i.e., $\|\bm{\beta}_{kc} - \bm{\beta}_{kc'}\|_1 > h$ for some large $h > 0$. 

We can regard the data collection procedure at each hospital as a sampling process from the total population. As inherited from the total population, all the hospitals will share the same set of subgroups. However, EHR data collected at each hospital is essentially a convenience sample, affected by factors such as location and administration patterns of the hospital.
As a result, due to the location and administration differences between hospitals,
each hospital may have a significantly different set of subgroup mixing proportions. Therefore, the observed non-negligible between-study heterogeneity can be reasonably well described by the hospital-specific mixing proportions, while there is little between-study heterogeneity within each subpopulation, i.e., $\|\bm{\beta}_{kc} - \bm{\beta}_{jc}\|_1 < h$ for some small $h > 0$. 
This motivates us to consider the following two-step information transfer approach.

Specifically, here we aim to investigate the predictive relationship between risk factors and PASC development, adjusted by patients' baseline health statuses. For each observation $(y_{ki}, \X_{ki}, \Z_{ki})$, the goal of our study is to investigate the conditional joint distribution  
\begin{align}\label{factorization}
    f_{y,\Z|\X}^k(y_{ki}, \Z_{ki}|\X_{ki} ) =  f_{y|\Z,\X}^k(y_{ki}| \Z_{ki},\X_{ki} )f_{\Z|\X}^k(\Z_{ki}|\X_{ki}).
\end{align}
The subpopulation structure of interest is the one that ensures $\Z_{ki}$ is independent of the joint distribution of $(y_{ki}, \X_{ki})$ when conditioned on that subpopulation structure. This conditional independence is essential for achieving the desired decomposition of both $f_{\Z|\X}^k(\Z_{ki}|\X_{ki})$ and $f_{y|\Z,\X}^k(y_{ki}| \Z_{ki},\X_{ki} )$, as will be elaborated below.

Our first step is to  use the information in $\{\Z_{ki}\}_{k,i}$  to identify and match the subpopulations across all the studies through a heterogeneity-aware finite mixture model 
\begin{align}
    f_{\Z|\X}^k(\Z_{ki}|\X_{ki}) = \sum_{c=1}^C P(\Z_{ki}|C_{ki}=c)P(C_{ki}=c) = \sum_{c=1}^C \lambda_{kc} f_{\Z}(\Z_{ki};\bpi_c). \label{decomp_z}
\end{align}
The first equality in \eqref{decomp_z} relies on the above conditional independence relationship, i.e.,  
$P(\Z_{ki}|C_{ki}=c,\X_{ki})=P(\Z_{ki}|C_{ki}=c)$, 
and the assumption that the predictors $\X_{ki}$ contain no information of the subpopulation structure.
As a result, in \eqref{decomp_z}, 
$f_{\Z}(\Z_{ki};\bpi_c)=P(\Z_{ki}|C_{ki}=c)$  characterizes the subpopulation-specific distribution of $\Z_{ki}$, and $\lambda_{kc}=P(C_{ki}=c)$ denotes the mixing proportion of the $c$-th subpopulation at the $k$-th study that satisfies $\sum_{c=1}^C \lambda_{kc} = 1$.
Note that, we let all studies share parameters $\bPi=(\bpi_1,\ldots,\bpi_C)$ to emphasize the idea of the shared subpopulations, and we allow $\lambda_{kc} \neq \lambda_{jc}$ for any $k\neq j$ across all $c=1,\ldots,C$ to take account the between-study heterogeneity. We collect subgroup mixing proportions across all studies into $\Lambda=(\Lambda_1,\ldots, \Lambda_K)$ with $\Lambda_k=(\lambda_{k1},\ldots,\lambda_{kC})^T$. The specific form of $f_{\Z}(\Z_{ki};\bpi_c)$ is flexible, and can be selected to satisfy different analysis demands. Let $\bm{\theta}_{str}=(\bm{\Pi}, \Lambda)$ denote all the parameters involved in subpopulation structure identification. After specifying the $f_{\Z}(\cdot)$ function and the number of subpopulations $C$, by fitting
\begin{align}
    L_{\Z}(\bm{\theta}_{str}) = \prod_{k=0}^K\prod_{i=1}^{n_k} \sum_{c=1}^C \lambda_{kc} f_{\Z}(\Z_{ki};\bpi_c) \label{fmm}
\end{align} 
to $\{\Z_{ki}\}_{k,i}$ observed from all studies, we can get the estimated $\widehat{\bm{\theta}}_{str}=(\widehat \bPi, \widehat \Lambda)$. The $\widehat{\bm{\theta}}_{str}$ can be used to calculate the initial subpopulation membership probability for each patient, whose accuracy will be further improved in the second step by incorporating the subpopulation structure information contained in $f_{y|\Z,\X}^k(y_{ki}| \Z_{ki},\X_{ki} )$.  


At the second step, the information transfer should be conducted within each subpopulation so that all the available source studies can participate in the analysis. 
We have
\begin{align*} 
    f_{y|\Z,\X}^k(y_{ki}| \Z_{ki},\X_{ki}) = \sum_{c=1}^C f(y_{ki}|\X_{ki}, C_{ki}=c; \bbeta_{kc})P(C_{ki}=c|\Z_{ki};\bm{\theta}_{str} )
\end{align*}
by the same assumptions we used for deriving \eqref{decomp_z}, 
which lead to $f(y_{ki}|\X_{ki}, \Z_{ki}, C_{ki} = c; \bbeta_{kc})=f(y_{ki}|\X_{ki}, C_{ki} = c; \bbeta_{kc})$ and 
$P(C_{ki}=c|\X_{ki}, \Z_{ki}; \bm{\theta}_{str})=P(C_{ki}=c|\Z_{ki}; \bm{\theta}_{str})$. 
This gives us
\begin{align}\label{ly}
   L_{y}(\bm{\theta}_{reg};\widehat{\bm{\theta}}_{str})&=\prod_{k=0}^K\prod_{i=1}^{n_k} \sum_{c=1}^C f(y_{ki}|\X_{ki}, C_{ki}=c;\bbeta_{kc})P(C_{ki}=c|\Z_{ki}; \widehat{\bm{\theta}}_{str}), 
\end{align} 
where each patient's initial subpopulation membership probability is $\widehat{v}_{kic}=P(C_{ki}=c|\Z_{ki}; \widehat{\bm{\theta}}_{str})$ by using the Bayes' rule and the subpopulation structure information learned at the first step. 
Based on \eqref{ly}, general transfer learning techniques \citep{li2020transfer, liang2020learning} can be applied within each subpopulation to borrow information from all source studies to improve the estimation of $\B_0$. However, it is noteworthy that in this setting, since transfer learning is conducted within each subpopulation derived from the first step, a significant research gap exists: existing theoretical analyses of transfer learning estimators do not account for the uncertainty introduced by the estimation of $\bm{\theta}_{str}$. This issue is addressed in detail in Section \ref{section:theory}.



Take the approach proposed in \citet{tian2022transfer} as an example, we first fit a working model by assuming $\B_0=\B_1=\ldots=\B_K=\B:=(\bm{\beta}_1, \ldots, \bm{\beta}_C)$ in \eqref{ly} to obtain a biased estimator of $\B_0$ with data from all studies, i.e.,
\begin{equation} \label{alg:optim1}
    \widehat{\B}=\argmin_{\B\in\mathbb{R}^{p\times C}}\left\{-\log L_{y}(\bm{\theta}_{reg};\widehat{\bm{\theta}}_{str})+\lambda_{pool}\|\B\|_1\right\} \quad \text{s.t.} \quad  \B_0=\B_1=\ldots=\B_K=\B, 
\end{equation}
where $\lambda_{pool} \geq 0$ is a tuning parameter. The $\ell_1$ penalty is applied here to achieve variable selection. Then, we perform a bias-correction to the biased estimator $\widehat{\B}$ using only the data from the target study to get the final estimator $\widehat{\B}_0 =\widehat{\B} + \widehat{\bm{\Delta}}$ with $\widehat{\bm{\Delta}}=(\widehat{\bm{\delta}}_{1},\ldots,\widehat{\bm{\delta}}_{C})$ obtained from
\begin{equation} \label{alg:optim2}
\widehat{\bm{\Delta}}=\argmin_{\bm{\Delta}\in\mathbb{R}^{p\times C}}\left\{ -\log\prod_{i=1}^{n_0} \sum_{c=1}^C \widehat{v}_{0ic} P(y_{0i}|\X_{0i}, C_{0i}=c;\widehat{\B} +\bm{\Delta}) + \lambda_{bias}\|\bm{\Delta}\|_1\right\},
\end{equation}
where $\lambda_{bias}\geq 0$ is a tuning parameter.

 
Note that both the optimization problems \eqref{alg:optim1} and \eqref{alg:optim2} can be solved by EM algorithm \citep{mclachlan2007algorithm}. Let's first consider the EM algorithm for obtaining the pooled biased estimator by solving \eqref{alg:optim1}, and the bias correction step  \eqref{alg:optim2} can be completed similarly. For simplicity, we write $f(y_{ki}|\X_{ki}, C_{ki}=c;\bbeta_{kc})$ as $f(y_{ki},\bbeta_{kc})$. By treating the subpopulation membership $C_{ki}$ as missing, based on the complete data, 
the working model obtained by assuming $\B_0=\dots=\B_K=\B$ can be written as  
\begin{align}\label{comp_llik}
    \sum_{k=0}^K\sum_{i=1}^{n_k}\sum_{c=1}^C I(C_{ki}=c) \log\{ f(y_{ki},\bbeta_{c})\}.
\end{align}
At the $t$-th iteration, denote the current estimator as $\B^t=(\bbeta_{1}^t,\ldots,\bbeta_{C}^t)$. We first perform $E$-step to construct the $Q$-function by taking expectation to \eqref{comp_llik} and get
\begin{align*}
    Q(\B|\B^t)=\sum_{k=0}^K\sum_{i=1}^{n_k}\sum_{c=1}^C P(C_{ki}=c|y_{ki},\X_{ki},\Z_{ki},\B^t, \widehat{\bm{\theta}}_{str}) \log\{f(y_{ki},\bbeta_{c})\},
\end{align*}
where 
\begin{align}\label{Estep}
P(C_{ki}=c|y_{ki},\X_{ki},\Z_{ki},\B^t, \widehat{\bm{\theta}}_{str})=\frac{\widehat{v}_{kic}f(y_{ki},\bbeta_{c}^t)}{\sum_{c=1}^C \widehat{v}_{kic}f(y_{ki},\bbeta_{c}^t)}\coloneqq w_{kic}^t
\end{align}
incorporates the regression information into the subpopulation membership probability estimation, which 
refines $\widehat{v}_{kic}=P(C_{ki}=c|\Z_{ki},\widehat{\bm{\theta}}_{str})$ estimated from the first step. When initialize this EM algorithm, we directly let $w_{kic}^0=\widehat{v}_{kic}$ and there is no need to specify $\bbeta_{c}^0$. The $M$-step is to maximize the $Q$-function with the penalization over $\B$. After ignoring the terms without unknown parameters, the $M$-step is 
\begin{align} \label{Mstep:exp}
    \B^{t+1} \leftarrow  \argmin_{\B\in\mathbb{R}^{p\times C}}\left\{ \sum_{k=0}^K\sum_{i=1}^{n_k}\sum_{c=1}^C w_{kic}^t  L_{k}(\bm{\beta}_{c}; y_{ki}, \bm{x}_{ki})+ \lambda_{pool} \|\B\|_1 
    \right\}
\end{align}
where $L_{k}(\bbeta_{c}; y_{ki}, \bm{x}_{ki})=-y_{ki} \bm{x}_{ki}^\top \bm{\beta}_{c} +g(\bm{x}_{ki}^\top \bm{\beta}_{c}).$
The penalty level at each subpopulation can be different, in which case we can assign different tuning parameters for each subgroup by replacing $\lambda_{pool} \|\B\|_1$ with $\sum_{c=1}^{C}\lambda_{pool,c} \|\bbeta_{c}\|_1$.  
This optimization problem can be solved by using R package `glmnet' \citep{hastie2021introduction}. 
By repeating the $E$-step and $M$-step until certain pre-specified convergence rules are met, we get the pooled estimator $\widehat{\B}$ from the working model. By similarly applying the EM algorithm to solve \eqref{alg:optim2}, we can get the bias correction term $\widehat{\bm{\Delta}}$ and derive the bias corrected estimator by $\widehat \B_0=\widehat \B +\widehat{\bm{\Delta}}$.  
The algorithm that implements the proposed method is summarized in Algorithm \ref{alg:2}, where the detailed $E$-step and $M$-step are presented. 

    \begin{algorithm}[!htb]
\caption{Targeted learning via probabilistic subpopulation matching}\label{alg:2}
\textbf{Input:} $\Z_{ki}, \X_{ki}, y_{ki},k=0,\dots,K, i=1,\dots,n_k$, the subpopulation number $C$, the threshold $\tau$ and the maximal iteration number $M$.

\begin{algorithmic}[1]
\STATE Fit a finite mixture model based on $L_z(\bm{\theta}_{str})$ to obtain the estimators $\widehat{\bm{\theta}}_{str}$.
\STATE \textbf{Joint estimation:}
\STATE \qquad \textbf{For} $t=1, \ldots, M$,
\STATE \qquad \qquad  \textbf{$E$-step:} Calculate the posterior weights, 
$$
w_{kic}^t=\begin{cases}
\widehat{v}_{kic} & \text{if} \quad t=1, \\
 \frac{\widehat{v}_{kic}f(y_{ki},\bbeta_{c}^t)}{\sum_{c=1}^C \widehat{v}_{kic}f(y_{ki},\bbeta_{c}^t)} & \text{if} \quad t>1.
\end{cases}
$$
\STATE \qquad \qquad \textbf{$M$-step:} Minimize the penalized weighted loss function
$$
   \B^{t+1} \leftarrow  \argmin_{\B\in\mathbb{R}^{p\times C}}\left\{ \sum_{k=0}^K\sum_{i=1}^{n_k}\sum_{c=1}^C w_{kic}^t  L_{k}(\bm{\beta}_{c}; y_{ki}, \bm{x}_{ki})+ \lambda_{pool} \|\B\|_1 
    \right\}
$$
\STATE \qquad \qquad \textbf{If} $\left\| \B^{t+1}-\B^{t}\right\|_2/\left\| \B^{t}\right\|_2\le \tau $, then let $t=M$ and break. 
\STATE Denote $\widehat{\B}=\B^{M}$.
\STATE \textbf{Correcting bias:}
\STATE \qquad \textbf{For} $t=1, \ldots, M$,
\STATE \qquad \qquad  \textbf{E-step:} Calculate the posterior weights, 
$$
w_{0ic}^t=\begin{cases}
\widehat{v}_{0ic} & \text{if} \quad t=1, \\
 \frac{\widehat{v}_{0ic}f(y_{0i},\widehat{\bbeta}_{c}+\bm{\delta}_{c}^t)}{\sum_{c=1}^C \widehat{v}_{0ic}f(y_{0i},\widehat{\bbeta}_{c}+\bm{\delta}_{c}^t)} & \text{if} \quad t>1.
\end{cases}
$$
\STATE \qquad \qquad \textbf{M-step:} Minimize the penalized weighted loss function
$$
   \bm{\Delta}^{t+1} \leftarrow  \argmin_{ \bm{\Delta}\in\mathbb{R}^{p\times C}}\left\{ \sum_{i=1}^{n_0}\sum_{c=1}^C w_{0ic}^t  L_{0}(\widehat{\bm{\beta}}_{c}+\bm{\delta}_{c}; y_{0i}, \bm{x}_{0i})+ \lambda_{bias} \|\bm{\Delta}\|_1 
    \right\}
$$
\STATE \qquad \qquad \textbf{If} $\left\| \bm{\Delta}^{t+1}-\bm{\Delta}^{t}\right\|_2/\left\| \bm{\Delta}^{t}\right\|_2\le \tau $, then let $t=M$ and break. 
\STATE Denote $\widehat{\bm{\Delta}}=\bm{\Delta}^{M}$.
\end{algorithmic}
\textbf{Output:} $\widehat{\B}_0=\widehat{\B}+\widehat{\bm{\Delta}}$.
\end{algorithm}

\begin{remark}\label{remark:6} 
If we execute the $E$-step and $M$-step in Algorithm \ref{alg:2} only once, i.e., setting $M=1$, this is equivalent to fitting a transfer learning model for each subpopulation separately using a weighted loss function with an $\ell_1$-norm penalty, where the weights $\hat{v}_{kic}$ are applied to each observation $(y_{ki}, \X_{ki})$. This is referred to as the one-step algorithm. However, this approach entirely ignores the subpopulation structure information that the regression $y_{ki}|(\X_{ki}, C_{ki}=c)$ can provide. In contrast, Algorithm \ref{alg:2} iteratively updates the weights $w_{kic}^t$ in the EM algorithm to incorporate the subpopulation structure information provided by $y_{ki}|(\X_{ki}, C_{ki}=c)$, leading to more accurate weight estimation. Consequently, the regression parameter estimation also benefits from this improved weight accuracy. 
We show this phenomenon in our numerical studies and demonstrate the superior performance of Algorithm \ref{alg:2} compared to the one-step algorithm.
\end{remark}

\begin{remark}
Our two-step estimation procedure is mainly motivated by the factorization of the likelihood function into the product of $L_{y}(\bm{\theta}_{reg};\bm{\theta}_{str})$ and $L_{\Z}(\bm{\theta}_{str})$, as implied by \eqref{factorization}. 
An alternative approach is to directly maximize the likelihood function with respect to all the unknown parameters $\bm{\theta}=(\bm{\theta}_{reg},\bm{\theta}_{str})$, which, however, could be ill-behaved and computationally complicated \citep{chen2010asymptotic}.
Our procedure essentially applies the pseudolikelihood method \citep{gong1981pseudo} by first estimating the structure parameters $\bm{\theta}_{str}$ and then maximizing $L_{y}(\bm{\theta}_{reg};\widehat{\bm{\theta}}_{str})$ with respect to the regression parameters by treating $\bm{\theta}_{str}$ as fixed at its estimated values. This approach results in a more stable and computationally efficient estimation process. 
\end{remark}


Based on the outputs of Algorithm \ref{alg:2} and the observation from a new patient $(\X_{0i}^*,\Z_{0i}^*)$ at the target study, the risk prediction can be obtained by a weighted average of the risk predictions made at all subpopulations. Taking the logistic regression as an example, we have the risk probability estimated by $P(y_{0i}^*=1|\X_{0i}^*,\Z_{0i}^*)=\sum_{c=1}^C P(y_{0i}^*=1|\X_{0i}^*, C_{0i}^*=c; \hat\bbeta_{0c})P(C_{0i}^*=c|\Z_{0i}^*; \widehat{\bm{\theta}}_{str}).$
When a proper threshold value of the risk is provided, a prediction can be made based on a comparison between the threshold value and the estimated risk probability. 


\subsection{Example: using latent class analysis to identify subpopulation structure}

In the PASC study, some binary variables are collected to show the absence or presence of certain chronic diseases before infection. 
These pre-infection chronic conditions can be used as $\Z_{ki}$ to categorize baseline health status for each patient. 
A commonly used tool to find distinct subgroups of a complex population based on the co-occurrence pattern of several diseases is the latent class analysis (LCA) \citep{bandeen1997latent}. We use LCA as an example of the finite mixture model to illustrate our proposed approach in PASC study. LCA uses the unobserved latent class structure to take account the correlation between the observed variables, which makes the specification of $f_{\Z}(\Z_{ki};\bpi_c)$ be straightforward.   Specifically, based on the binary variables $\Z_{ki}=(z_{ki1},\ldots,z_{kiq})^T\in \{0,1\}^q$, we have $f_{\z}^k(\Z_{ki}) = \sum_{c=1}^C \lambda_{kc} f_{\Z}(\z_{ki};\bpi_c)=\sum_{c=1}^C \lambda_{kc} \prod_{j=1}^q \pi_{cj}^{z_{kij}}(1-\pi_{cj})^{1-z_{kij}}$ by assuming that the binary variables $z_{kij}$s are independent to each other within each latent class. For each subpopulation, we have its specific parameter $\bpi_c=(\pi_{c1}, \ldots, \pi_{cq})^T$ where $\pi_{cj}$ denotes the pravelance of the $j$-th disease for patients within the $c$-th subpopulation.

{

}

\section{Theoretical Results}\label{section:theory}

In this section, we establish theoretical guarantees for Algorithm \ref{alg:2}. By incorporating a subpopulation structure into the analysis, the between-study heterogeneity within each subpopulation can be controlled to a small value, facilitating knowledge transfer. However, this approach also introduces uncertainty in the estimation of $\bm{\theta}_{str}$ at the first step. We focus on investigating how these factors affect the estimation of $\B_0$.

We need the following regularity conditions.
\begin{itemize}
	\item[A1.]  For all $k=0, 1, \ldots, K$ and any $\bm{\beta} \in \mathbb{R}^p $, $\bm{x}_{ki}^{\top} \bm{\beta}$'s are independently distributed sub-Gaussian variables with mean zero and $\mathbb{E}(\bm{x}_{ki}^{\top} \bm{\beta})^2\leq \kappa \| \bm{\beta} \|_2^2$, where $\kappa$ is a positive constant. Denote the covariance matrix of $\bm{x}_{ki}$ as $\bm{\Sigma}_{k}$. We have $0< \kappa_l \le \inf_k \lambda_{\min} (\bm{\Sigma}_{k}) \le \sup_{k} \lambda_{\max}(\bm{\Sigma}_{k}) \le \kappa_u <\infty$, where $\kappa_l$ and $\kappa_u$ are both  positive constants.
	
	\item[A2.]  The function $g$ is infinitely differentiable and strictly convex, i.e., its second derivative satisfies $g^{\prime\prime}(x)\ge m_g>0$ with $m_g$ a positive constant. 

   \item[A3.]  $\sup \limits_{k,i} \|\bm{x}_{ki} \|_{\infty} \le M_x < \infty$ almost surely and 
	 $\sup \limits_{k, c} \sup \limits_{|\epsilon|<U}  | g^{\prime\prime}( \bm{x}_{ki}^{\top} \bm{\beta}_{kc}+\epsilon)| \le M_g < \infty$ almost surely, where $U$, $\epsilon$, $M_x$ and $M_g$ are some positive constants.

\end{itemize}

Assumption A1 requires independent sub-Gaussian design with  well-behaved covariance matrices. The covariance matrix $\bm{\Sigma}_k$ may vary across different studies, making it more suitable for real-world data applications. Assumption A2 imposes the differentiability and strict convexity  of $g(\cdot)$.   Assumption A3 requires that  the second-order  derivative $g^{\prime\prime}(x)$ is upper bounded within a bounded region. Assumptions A1, A2 and A3 are commonly used in the study of high-dimensional GLMs and hold for linear, logistic and Poisson regression models, see \citet{negahban2012unified,li2023estimation} and reference therein for more information.  {In addition to the standard assumptions for GLMs, to show the reduced between-study heterogeneity after accounting for the subpopulation structure, 
we require that the true parameter $\bm{\beta}_{kc}$ at the $k$-th source study 
be sufficiently close to  $\bm{\beta}_{0c}$. Specifically, we consider the following parameter space
$$
\bm{\Theta}(s,h)=\left\{\B_k\in\mathbb{R}^{p\times C}, k=0,\ldots,K: \sup\limits_{c} \|\bm{\beta}_{0c} \|_0\le s,   \sup\limits_{k, c} \|\bm{\beta}_{0c}-\bm{\beta}_{kc} \|_1\le h \right\}
$$
where $s$ is a positive integer to characterize a sparse model, and $h$ is  a small positive constant that measures the similarity between the source studies and the target study.} Moreover, we need the estimator $\widehat{\bm{\theta}}_{str}$ obtained from the first step of finite mixture model fitting be close enough to  $\bm{\theta}_{str}$. Note that, for $c \in \{1, \ldots, C\}$, the joint estimation step under class $c$ can be understood as solving the following equation 
\begin{equation}  
	\sum_{k=0}^{K}\left[ \sum_{i=1}^{n_k} w_{kic} \left( -y_{ki} \bm{x}_{ki}^\top +g^{\prime} (\bm{x}_{ki}^\top  \bm{\beta}_c)\bm{x}_{ki}^\top \right)\right]=0, \notag
\end{equation}
where $\bm{\beta}_{c}$ is the regression parameter at the $c$-th subpopulation in the working model, and this equation converges to its population version 
\begin{equation} \label{eqtrans}
\sum_{k=0}^{K}\alpha_k\mathbb{E}\left[ \sum_{i=1}^{n_k} w_{kic} \left( g^{\prime} (\bm{x}_{ki}^\top  \bm{\beta}_c) -g^{\prime} (\bm{x}_{ki}^\top  \bm{\beta}_{kc})\right)\bm{x}_{ki}^\top\right]=0, 
\end{equation}
where $\alpha_k=n_k/(N+n_0)$ and $N=\sum_{k=1}^K n_k$ is the total sample size of all source studies. We also require that $\bm{\beta}_{c}$ is close to $\bm{\beta}_{0c}$. Therefore, we need the following assumptions:

\begin{itemize}
	\item[A4.]  Assume that the parameters $\widehat{\bm{\theta}}_{str}$ estimated from the finite mixture model satisfy $\|\widehat{\bPi}-\bPi \|_2 \vee \|\widehat{\Lambda} -\Lambda \|_2 \le R_{N} $, {and that the initial subpopulation membership probability  $v_{kic}=P(C_{ki}=c|\Z_{ki}; \bm{\theta}_{str})$ satisfies $v_0< v_{kic} < 1-v_0$ for a small constant $v_0$ and all $k=0,\ldots, K$, $i=1, \ldots, n_k$, $c=1, \ldots, C$.}
	\item[A5.]  Denote $\widetilde{\bm{\Sigma}}_{c}^{h}=\sum_{k=0}^{K} \alpha_k \mathbb{E} \left[ \frac{1}{n_k}\sum_{i=1}^{n_k}w_{kic} \left(\int_{0}^{1}g^{\prime\prime}(\bm{x}_{ki}^{\top}\bm{\beta}_{0c}+t \bm{x}_{ki}^{\top}(\bm{\beta}_{c}-\bm{\beta}_{0c}) )dt \right) \bm{x}_{ki}\bm{x}_{ki}^{\top}  \right]$ and $\widetilde{\bm{\Sigma}}_{kc}^{h}= \mathbb{E} \left[ \frac{1}{n_k}\sum_{i=1}^{n_k}w_{kic}\left(\int_{0}^{1}g^{\prime\prime}(\bm{x}_{ki}^{\top}\bm{\beta}_{0c}+t \bm{x}_{ki}^{\top}(\bm{\beta}_{kc}-\bm{\beta}_{0c}) )dt \right) \bm{x}_{ki}\bm{x}_{ki}^{\top}  \right]$.  We assume that  $ \| (\widetilde{\bm{\Sigma}}_{c}^{h})^{-1} \widetilde{\bm{\Sigma}}_{kc}^{h} \|_1 < \infty$ holds for all $k=0, \ldots,$ $K$ and $c=1, \ldots, C$.
\end{itemize}

Assumption A4 measures the estimation error of $\widehat{\bm{\theta}}_{str}$, 
 and its rate can achieve $O_p(1/\sqrt{N+n_0})$
 since the finite mixture model fitting is based on data from all studies. {It also assumes that each sample has a positive probability of coming from any of these $C$ subpopulations.} 
 Assumption A5 is similar to Assumption 4 in \citet{tian2022transfer}, and it ensures that the working model parameter $\bm{\beta}_c$ in \eqref{eqtrans} is close to the target parameter $\bm{\beta}_{0c}$, such that $\|\bm{\beta}_{0c}-\bm{\beta}_{c} \|_1 \le C_{\Sigma} h$, where $C_{\Sigma} $ is a positive constant.

Without loss of generality, we focus on the case where the number of subpopulations is two, i.e., $C=2$, to derive our theoretical results. The extension to the situation where $C > 2$ is straightforward. 
{The joint estimation step involves an EM-algorithm, whose convergence analysis requires properly defined signal strength and contraction region.
Let $\delta= \| \bm{\beta}_{1}-\bm{\beta}_{2} \|_2$ to measure the signal strength in the considered subpopulation-based working model.} With a positive constant $0<c_{\delta}<1/2$, the contraction basin is defined as
	$\mathcal{B}(\bm{\beta}_{1},\bm{\beta}_{2}, c_{\delta}, \delta )= \left\{ (\check{\bm{\beta}}_{1}, \check{\bm{\beta}}_{2}): \|\check{\bm{\beta}}_{1}- \bm{\beta}_{1} \|_{2} \vee \|\check{\bm{\beta}}_2- \bm{\beta}_{2} \|_2  \le c_{\delta}\delta \right\}.$

To guarantee convergence, the signal strength $\delta$ should be large enough to satisfy $\delta> C(c_0, v_0, \kappa_l, \kappa_u, m_g, M_g, s, h)$, which is a large constant related to the quantities within the parentheses.
{
Intuitively, the constant $C(c_0, v_0, \kappa_l, \kappa_u, m_g, M_g, s, h)$ measures the difficulty to recover the underlying model and also depends on the properties of the underlying model. 
For example, it relates to the covariance matrix $\bm{\Sigma}_k$ through the 
eigenvalues $\kappa_l$ and $\kappa_u$: the value of $C(c_0, v_0, \kappa_l, \kappa_u, m_g, M_g, s, h)$ decreases as the ratio $\kappa_l/\kappa_u$ increases. 
The properties of the considered GLM, e.g., the second derivative $g^{\prime\prime}(x)$, also affect its magnitude. 
Moreover, the sparsity level $s$ and the similarity measurement $h$ between the studies also decide the difficulty of the problem, {where larger values of $s$ and $h$ lead to a larger $C(c_0, v_0, \kappa_l, \kappa_u, m_g, M_g, s, h)$}.  
The specific form of $C(c_0, v_0, \kappa_l, \kappa_u, m_g, M_g, s, h)$ is determined by equations (S.28) and (S.29) in Section S.1.3 of the Supplementary Material. 
The requirement $\delta> C(c_0, v_0, \kappa_l, \kappa_u, m_g, M_g, s, h)$ implies that the convergence of EM algorithm in the joint estimation step requires well-separated subpopulations, i.e., the distance between $\bm{\beta}_{1}$ and $\bm{\beta}_{2}$ should be large enough.}  
We first provide the estimation error bound for $\widehat{\bm{\beta}}_{c}$ in Theorem \ref{pro1}.


\begin{theorem} \label{pro1}
	(Convergence rate of $\widehat{\B}$ in the joint estimation step).  Suppose that Assumptions A1-A5 hold.  Assume $\delta> C(c_0, v_0, \kappa_l, \kappa_u, m_g, M_g, s, h)$, $N\gg n_0$, and $s\log p/(N+n_0)\le c_2$ for a small constant $c_2$. Let $\lambda_{pool}=c_{pool}\sqrt{\log p/(N+n_0)}$, where $c_{pool}$ is a large positive constant. Then, we have
	\begin{align*}
\|\widehat{\bm{\beta}}_{c}-\bm{\beta}_{c}  \|_2^2 &\lesssim  s\frac{\log p}{N+n_0}+ \{(s R_N)\vee h\} \left(R_N + \sqrt{\frac{\log p}{N+n_0}}\right)
\end{align*}	
 holds for $c=1, 2$ with probability at least  $1-N^{-1}$ when the iteration number $M$ is sufficiently large.
\end{theorem}

The first term $s\log p/(N+n_0)$ in Theorem \ref{pro1} 
is the typical Lasso $\ell_2$-type error bound when data from all studies are pooled together under the assumption of a homogeneous model. This represents the optimal error rate achievable in our setting. The second term in Theorem \ref{pro1} measures how the estimation error $R_N$ from the finite mixture model fitting and the similarity measure $h$ influence the estimation error of $\widehat{\B}$. 
When $R_N$ takes its best achievable rate $O_p(1/\sqrt{N+n_0})$,
if $h\ll s\sqrt{\frac{\log p}{N+n_0}}$ is small, then we can achieve the optimal rate $\frac{s\log p}{N+n_0}$; when $h> s\sqrt{\frac{\log p}{N+n_0}}$, then the error bound is dominated by $h\sqrt{\frac{\log p}{N+n_0}}$. 

The bias correction step involves another EM algorithm to estimate the bias correction term, for which we need define another contraction basin $\mathcal{B}(\bm{\beta}_{01}, \bm{\beta}_{02}, c_{\delta_0}, \delta_0)$ centered at $\bm{\beta}_{01}$ and $\bm{\beta}_{02}$ with $ \delta_0=\|\bm{\beta}_{01}-\bm{\beta}_{02} \|_2$.
To guarantee convergence in the bias correction step, the separation  $\delta_0$ between the two subpopulations should also be large enough to satisfy $\delta_0 > C_0(c_0, v_0, \kappa_l, \kappa_u, m_g, M_g, s, h)$. {The specific form of $C_0(c_0, v_0, \kappa_l, \kappa_u, m_g, M_g, s, h)$ is decided by (S.37) and (S.38) in Section S.2.2 of the Supplementary Material, from which we can also find that $C_0(c_0, v_0, \kappa_l, \kappa_u, m_g, M_g, s, h)$ should be larger than $C(c_0, v_0, \kappa_l, \kappa_u, m_g, M_g, s, h)$. This is the consequence of employing a two-step procedure, i.e., joint estimation and bias correction, in transfer learning, and it aligns well with our intuition. Specifically, the bias correction step is based on the estimators obtained from the joint estimation step, thereby the uncertainty in estimating $\bm{\beta}_{c}$ should also be taken account when conducting bias correction. As a result, in order to accurately recover the subpopulation-specific regression structure in the target study, the distance between $\bm{\beta}_{01}$ and $\bm{\beta}_{02}$ should be larger than the random error accumulated in two steps, leading to a larger $C_0(c_0, v_0, \kappa_l, \kappa_u, m_g, M_g, s, h)$ than $C(c_0, v_0, \kappa_l, \kappa_u, m_g, M_g, s, h)$.}
We summarize the theoretical guarantees for the estimator $\widehat{\B}_{0}$ 
in Theorem \ref{Thm1}.

\begin{theorem} \label{Thm1}
	(Convergence rate of the final estimator $\widehat{\B}_{0}$).  Suppose that Assumptions A1-A5 hold. Assume $\delta_0 > C_0(c_0, v_0, \kappa_l, \kappa_u, m_g, M_g, s, h)$ and $ s\sqrt{\frac{\log p}{n_0}} \le c_3$ for some small constant $c_3$. Let $\lambda_{pool}=c_{pool}\sqrt{\frac{\log p}{N+n_0}}$ and $\lambda_{bias}=c_{bias}\sqrt{\frac{\log p}{n_0}}$, where $c_{pool}$ and $c_{bias}$ are some positive constants that are large enough. Then, we have
	\begin{align*}
\|\widehat{\bm{\beta}}_{0c}-\bm{\beta}_{0c}  \|_2^2 &\lesssim  s\frac{\log p}{N+n_0}+ ((s R_N)\vee h) \left(R_N + \sqrt{\frac{\log p}{N+n_0}}\right)+\left( h\sqrt{\frac{\log p}{n_0}}+hR_{N} \right) \wedge (h^2)
\end{align*}	
 holds for $c=1, 2$ with probability at least  $1-n_0^{-1}$ when the iteration number $M$ is sufficiently large.
\end{theorem}

The first two terms of the above estimation error bound originate from the estimation error in the joint estimation step. The first term, $\frac{s\log p}{N+n_0}$, represents the optimal achievable error rate when applying Algorithm \ref{alg:2}, while the second term reflects how $R_N$ and $h$ influence the estimation of $\bm{\beta}_{c}$. The third term arises from the bias correction step, which also depends on $R_N$ and $h$.
Similarly, if the error $R_N$ incurred from the finite mixture model fitting is $O_p(1/\sqrt{N+n_0})$, then when $h\ll s\sqrt{\frac{\log p}{N+n_0}}$ is small, we can achieve the optimal rate $\frac{s\log p}{N+n_0}$; when  $ s\sqrt{\frac{\log p}{N+n_0}}< h \le \sqrt{\frac{\log p}{n_0}}$,  the error bound is dominated by $h^2$; while when $h>\sqrt{\frac{\log p}{n_0}}$, the error bound is dominated by $h\sqrt{\frac{\log p}{n_0}}$.

\begin{remark}
	In contrast to the classical transfer learning framework proposed by \citet{tian2022transfer} and \citet{li2023estimation}, the upper bound of the estimation error of our proposed estimator introduces new terms, namely $sR_{N}\vee h$ and $R_{N} h$, where $R_{N}$ represents uncertainties arising from the finite mixture model fitting, representing the cost of incorporating and estimating a subpopulation structure in the analysis.  
 If the estimation error $R_{N}$ is well controlled in the finite mixture model fitting, such that $R_N=O_p(1/\sqrt{N+n_0})$, then the error bound is reduced to $s\frac{\log p}{N+n_0}+h\sqrt{\frac{\log p}{n_0}}\wedge h^2 $, which matches the results in \citet{tian2022transfer}. 
 Furthermore, when $h \ll s\sqrt{\frac{\log p}{n_0}}$, the upper bound of the estimation error for the proposed estimator is better than the classical Lasso estimator error bound of $O_p(\frac{s\log p}{n_0})$, which is achieved using only the data from the target study.
\end{remark}

\section{Numerical Results}

In  this section, we illustrate the performance of the proposed method through simulation studies. We consider binary outcomes and use logistic regression in the simulation. We infer the subpopulation structure from several binary variables and the latent class analysis (LCA) is used whenever the subpopulation structure identification is needed. The methods under comparison include:
\begin{itemize}
    \item Targeted learning via probabilistic subpopulation matching approach (Targeted-PSM): our proposed approach that follows Algorithm \ref{alg:2} where each subpopulation has a specific tuning parameter.
    \item Targeted-PSM-1: the one-step implementation of Targeted-PSM by setting the iteration number to $M=1$. 
    \item LCA-GLM: a method that utilizes only the data from the target study to identify the subpopulation structure via LCA and then fit GLM prediction models. This method is similar to Targeted-PSM, but does not transfer knowledge from source studies.  
    \item Trans-GLM: a transfer learning method proposed by  \citet{tian2022transfer} that ignores the potential subpopulation structure.
    \item Naive-Lasso: a direct application of lasso regression to the target study. This approach disregards both the knowledge from source studies and the potential subpopulation structure in the data.
\end{itemize}

The target study sample size is $n_0=1500$, and the source study sample size is $n_k=1000$ for each $k\ge 1$. We increase the number of sites $K$ to evaluate the performance of different methods as a function of $K$. For the subpopulation structure, we consider three subpopulations $C=3$ and generate $q=5$ binary variables for latent class analysis under two settings. In the first setting, we let $\pi_{cj} \in \{0.1, 0.5, 0.9\}$ for $c=1, 2, 3$ and $j=1, \ldots, 5$, resulting in well-separated subpopulations as shown in Supplementary Table S1. 
In the second setting, we substitute 0.1, 0.5, and 0.9 with 0.3, 0.5, and 0.7, respectively, in Supplementary Table S1, to get less-separated subpopulations. When generating the mixing proportions of subpopulations in each study, we also consider two settings to change the level of heterogeneity between the target study and source studies. In one setting, the mixing proportions of the three subpopulations are similar between the target study and source studies, as shown in Supplementary Table S2, and in the other setting, we make the mixing proportions be significantly different, as shown in Supplementary Table S3.

For the covariates, we set the dimension to $p=100$ and generate  $\bm{x}_{ki} \overset{i.i.d.}{\sim} N_p(\mathbf{0}, \bm{\Sigma})$ for $k=0,\ldots,K$ and $i=1,\ldots,n_k$ where $\bm{\Sigma}=(\sigma_{ij})_{p\times p}$ with $\sigma_{ij}=0.5^{|i-j|}$. For the target study, the regression coefficient is set to be $\mathbf{B}_0=(\bbeta_{01}, \bbeta_{02}, \bbeta_{03})$ with $\bbeta_{0c}=0.5\cdot(\mathbf{1}_{s_c}, \mathbf{0}_{p-s_c})^{T}$ for $c=1, 2, 3$, where $\mathbf{1}_{s_c}$ contains $s_c$ ones and $ \mathbf{0}_{p-s_c}$ contains $p-s_c$ zeros. The 
support size 
of each $\bbeta_{0c}$ is set to $s_1=1$, $s_2=2$ and $s_3=6$. For the $k$-th source study, the regression coefficient matrix is  $\mathbf{B}_{k}=\mathbf{B}_0+(h/p) \mathcal{S}_{k}$, where $h \in \{5, 15\}$ and  $\mathcal{S}_{k}$ is a $p\times C$ sign matrix with its elements independently taking values from $\{-1, 1\}$ with equal probability 1/2. The responses are generated from a logistic regression model. We consider eight different settings and perform 100 replications under each setting with a correctly specified $C$. 

To evaluate the performance of various methods, we use mean square error (MSE) to assess the accuracy of the regression coefficient matrix $\B_0$ estimation. To demonstrate the performance of the proposed method in terms of prediction at the target study, we generate an additional 1500 samples from the same target data population as a testing set. We measure prediction accuracy on the testing set using the area under the receiver operating characteristic curve (AUC). Note that since Trans-GLM and naive-Lasso do not account for subpopulations, estimation accuracy is not evaluated for these two methods. 

\begin{figure}[!t]
    \centering
    \includegraphics[width=0.8\textwidth]{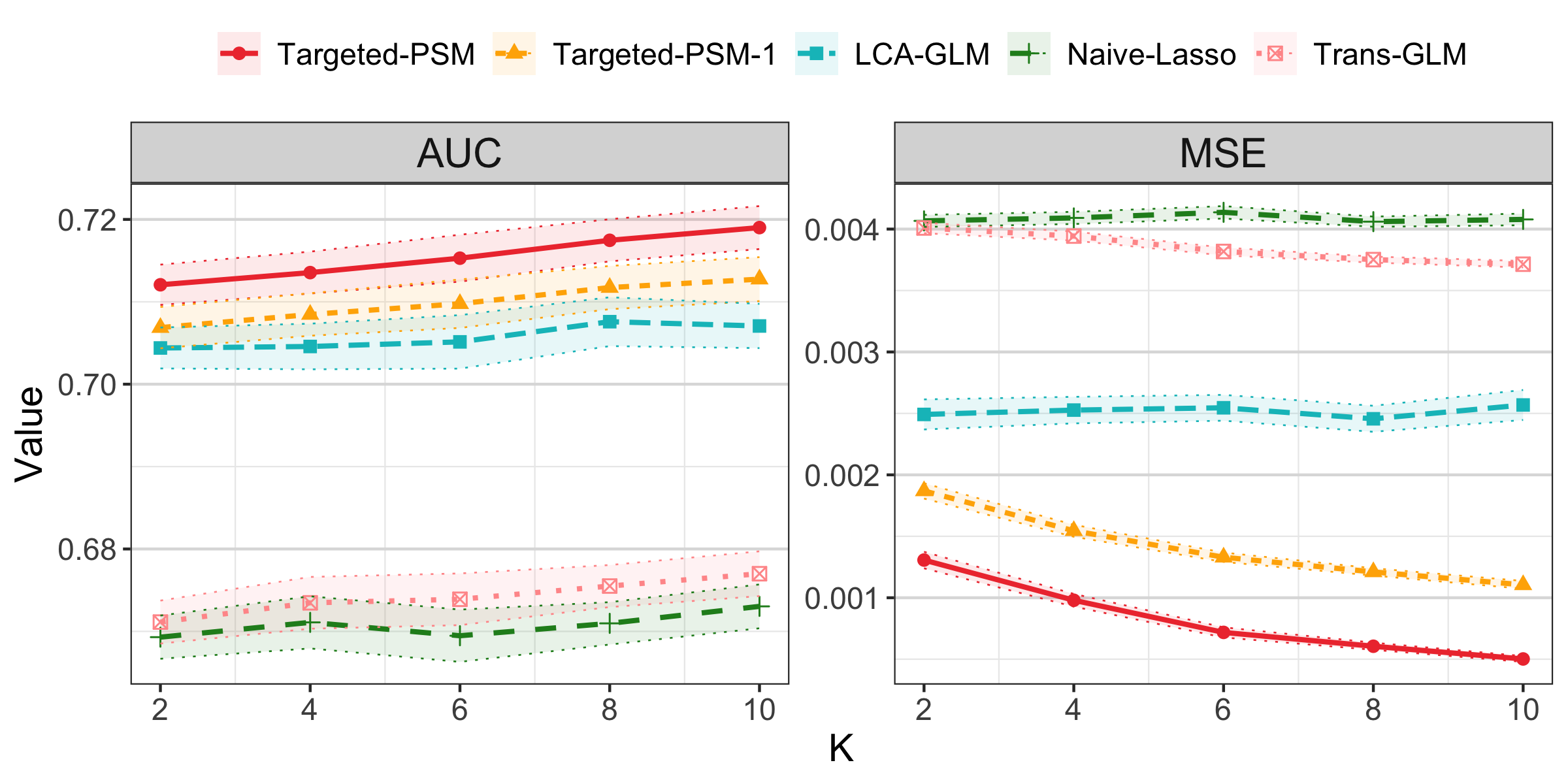}
    \caption{The estimation and prediction results for different methods when the subpopulations are well separated. 
    We let $h=5$ and the difference in mixing proportions between the target study and source studies be small in this setting.}
    \label{fig:1}
\end{figure}

\begin{figure}[!t]
    \centering
    \includegraphics[width=0.8\textwidth]{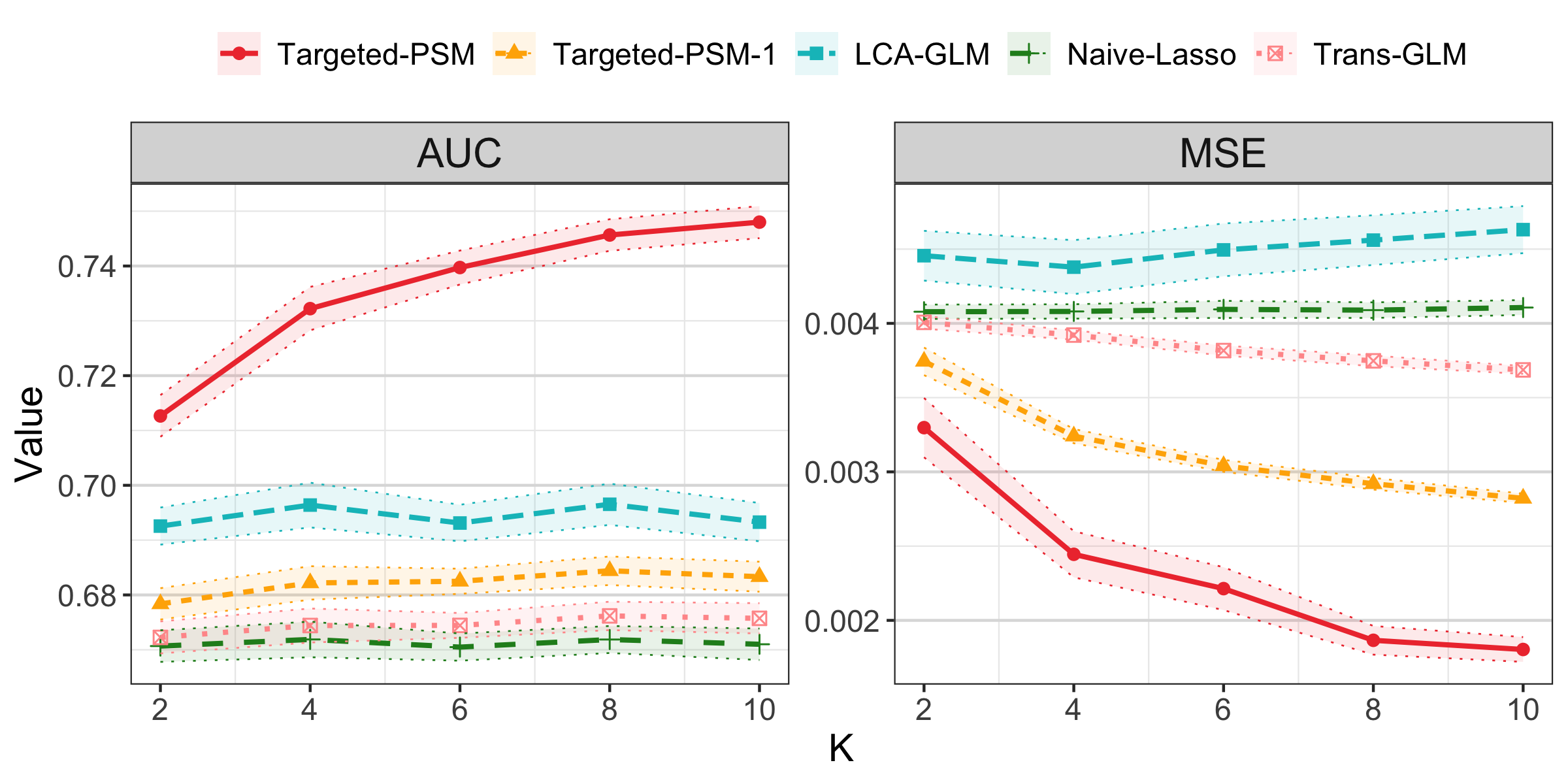}
    \caption{The estimation and prediction results for different methods when the subpopulations are not well separated. 
    We let $h=5$ and the difference in mixing proportions between the target study and source studies be small in this setting.}
    \label{fig:2}
\end{figure}

\begin{figure}[!ht]
    \centering
    \includegraphics[width=0.8\textwidth]{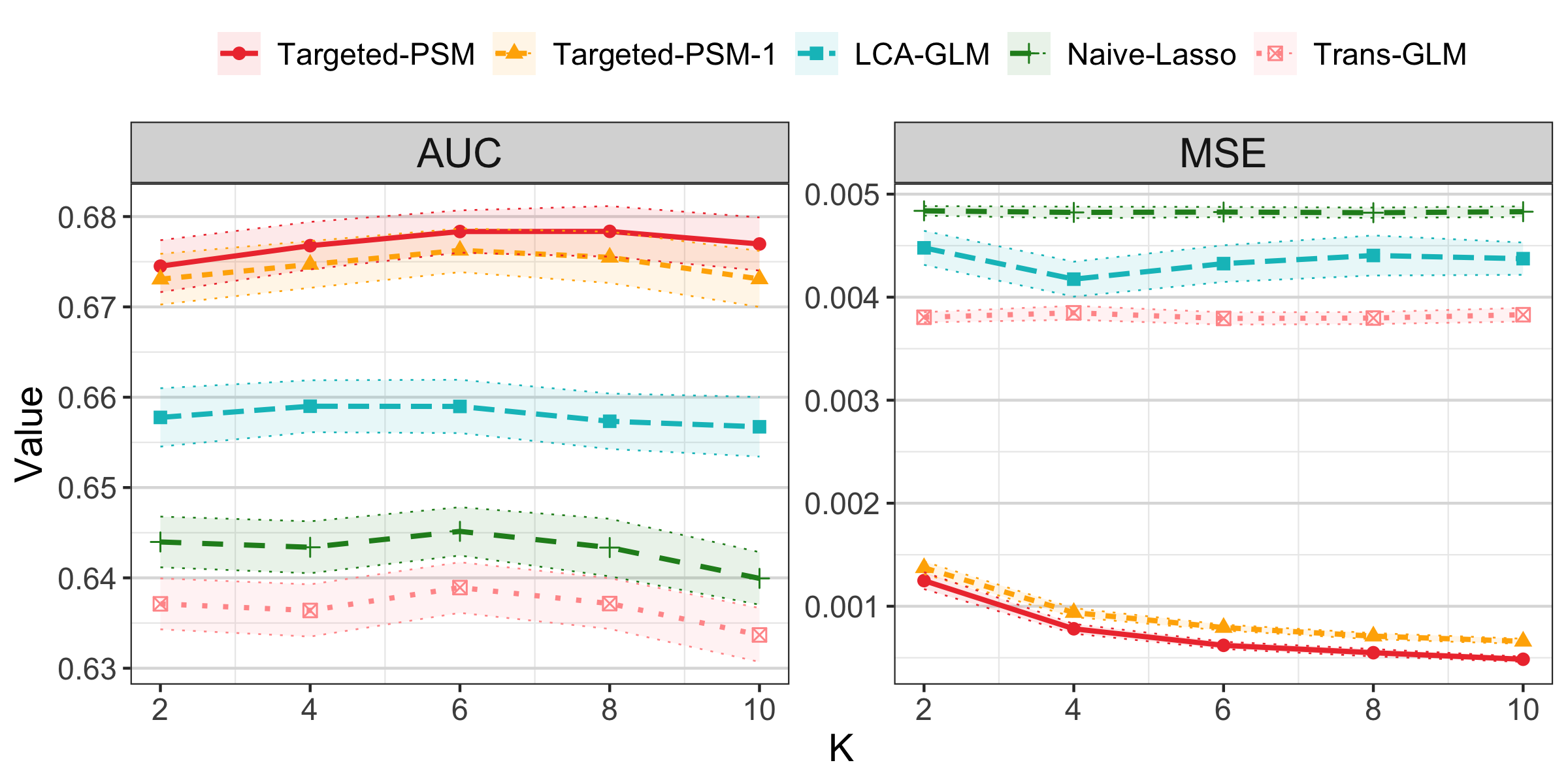}
    \caption{The estimation and prediction results for different methods when the subpopulations are well separated. 
    We let $h=5$ and the difference in mixing proportions between the target study and source studies be large in this setting.}
    \label{fig:5}
\end{figure}

\begin{figure}[!ht]
	\centering
	\includegraphics[width=0.8\textwidth]{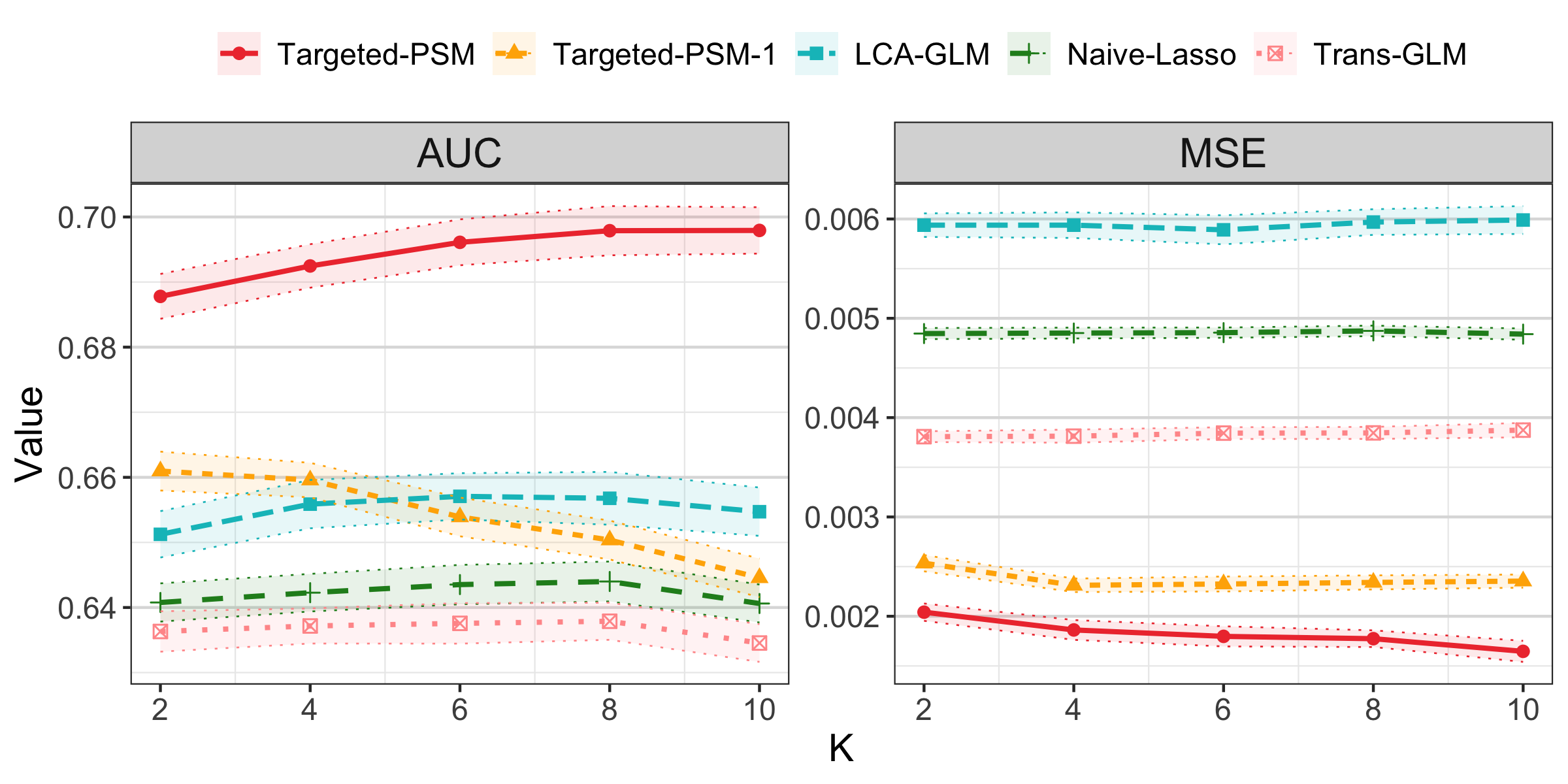}
	\caption{The estimation and prediction results for different methods when the subpopulations are not well separated. 
    We let $h=5$ and the difference in mixing proportions between the target study and source studies be large in this setting.}
	\label{fig:6}
\end{figure}

The simulation results for the settings with $h=5$ are presented in Figures \ref{fig:1}--\ref{fig:6}, with each dot in the figures representing the average of 100 replicates. In terms of estimation accuracy, across all settings, Targeted-PSM consistently achieves the lowest MSE, followed by Targeted-PSM-1. This highlights the importance of knowledge transfer and subpopulation membership probability refinement in improving estimation accuracy for the target study. As the number of source studies $K$ increases, MSE monotonically decreases for Targeted-PSM, showing the strength of this method to leverage information from source studies effectively regardless of the between-study heterogeneity. These two methods consistently outperform LCA-GLM, which only uses the target study data. The estimation performance of LCA-GLM does not change as $K$ increases, and it is highly dependent on the specific settings.  
Specifically, when the subpopulations are well-separated and distributed relatively evenly in the target study (see Figure \ref{fig:1}), LCA-GLM achieves its best estimation performance among the four settings. Moreover, all methods perform worse when the subpopulations are not well-separated.  


In terms of prediction accuracy, Targeted-PSM consistently outperforms all other methods across all scenarios, demonstrating its superior performance in prediction, with the AUC improving as $K$ increases. When the subpopulations are well-separated, Targeted-PSM-1 shows the second-best performance. However, when the subpopulations are less-separated, Targeted-PSM-1 performs no better than the local approach LCA-GLM. This further underscores the advantage of incorporating regression information to refine the subpopulation membership probabilities. Naive-Lasso and Trans-GLM both perform worse than the other three methods. 
The relative performance of Trans-GLM compared to Naive-Lasso depends on the level of between-study heterogeneity, as adjusted by the mixing proportions. 
Specifically, when the differences in the mixing proportions between the target study and source studies are small,  Trans-GLM outperforms Naive-Lasso. However, 
when there is a significant difference in mixing proportions, Trans-GLM may not perform as well as Naive-Lasso, suggesting that it might be less effective in targeted learning under conditions of substantial between-study heterogeneity.


The results for the settings with $h=15$ are displayed in Supplementary Figures S1 -- S4. As expected, except for the two methods that only rely on the target study information, i.e., LCA-GLM and Naive-Lasso, all other three methods get worse estimation and prediction results compared to the settings with $h=5$. The pattern observed in the method comparison, in terms of both estimation and prediction, remains consistent with the settings where $h=5$. Specifically, Targeted-PSM continues to show the best performance. 
For Trans-GLM, however, as $h$ increases and between-study heterogeneity grows, its performance declines across all four settings. It does not perform as well as the Naive-Lasso approach under these conditions, suggesting that Trans-GLM faces challenges in mitigating negative transfer effects in the scenarios considered.

\section{Discussion}


To fully utilize samples from all source studies for enhanced targeted learning while protecting the resulting 
model from negative transfers, we propose a probabilistic subpopulation matching approach. 
The key idea is to decompose the large between-study heterogeneity via a subpopulation structure and focus on subpopulation-level matching and knowledge transfer between studies. 
Our approach is devised as a two-step procedure: first, the subpopulation structure is identified to link studies at the subpopulation level; second, knowledge is transferred from all source studies to the target study within each subpopulation. 
This approach can effectively decompose both within- and between-study heterogeneity.
Although we illustrate our approach using finite mixture models in the first step, conceptually, any type of clustering method, such as k-means clustering, can be employed, making our approach more flexible. 
Incorporating the subpopulation structure allows our approach to move beyond the classical study-level matching framework. As its benefits, firstly, unlike most existing methods, our approach is simple to implement as it does not require finding an informative set. Additionally, there is minimal information loss since our approach enables every patient from any source study to contribute to the analysis, thereby achieving full utilization of the available data.

Our two-step estimation procedure is mainly motivated by the factorization of the joint likelihood function into the product of $L_{\Z}(\bm{\theta}_{str})$ and $L_{y}(\bm{\theta}_{reg};\bm{\theta}_{str})$, 
followed by the application of the pseudolikelihood approach as discussed in \citet{gong1981pseudo}. The pseudolikelihood approach is widely used to address computational issues caused by the ill-behaved joint likelihood, and it also offers desirable theoretical properties. Specifically, when the structure parameter estimator $\widehat{\bm{\theta}}_{str}$ is consistent and asymptotically normal, then under certain regularity conditions, the pseudolikelihood estimator $\widehat{\bm{\theta}}_{reg}$ can also be proved to be consistent and asymptotically normal \citep{gong1981pseudo, chen2010asymptotic}. However, these asymptotic properties of the pseudolikelihood estimator do not explicitly clarify how the uncertainty in $\widehat{\bm{\theta}}_{str}$ estimation impacts the estimation of $\bm{\theta}_{reg}$. 
To address this, we developed the non-asymptotic properties of $\widehat{\bm{\theta}}_{reg}$, where the estimation error of $\widehat{\bm{\theta}}_{str}$, i.e., $R_N$, appears as a component in the estimation error bound of $\widehat{\bm{\theta}}_{reg}$. Specifically, we prove that when the estimation error $R_N$ 
is sufficiently small, e.g., $R_N=O_p(1/\sqrt{N+n_0})$, and the subpopulation structure effectively reduces the between-study heterogeneity to be $h \ll s\sqrt{\log p/n_0}$, then the upper bound of the estimation error of the proposed method will be smaller than the classical Lasso $\ell_2$-type error bound $O_p(\frac{s\log p}{n_0})$  obtained when using only the target study data. Furthermore, if the subpopulation structure can reduce the between-study heterogeneity to satisfy $h\ll s\sqrt{\frac{\log p}{N+n_0}}$, then we can achieve the optimal rate $\frac{s\log p}{N+n_0}$ in transfer learning. These highlight the benefits of applying our method from a theoretical point of view.

From a practical point of view, our subpopulation-aware transfer learning approach is particularly useful in biomedical studies where diseases exhibit distinct progression paths across different subtypes. For instance, in the PASC study, tailoring predictions based on a patient's baseline health status allows for personalized guidance, enabling patients to take appropriate actions according to their specific health profiles. 
Beyond the PASC study, our approach is also highly relevant in cancer research, given the heterogeneous nature of many cancers whose distinct subtypes could differ in biological behavior, response to treatment, and prognosis \citep{breslow1986statistical}. 
Identifying and characterizing these subtypes and developing subtype-specific predictive models can facilitate more accurate prognostic assessments and tailor treatment strategies to individual patients, thereby improving therapeutic efficacy and reducing adverse effects. 
As another example, our approach is also well-suited for studying Alzheimer's disease, a neurodegenerative disease with distinct subtypes that manifest heterogeneous etiologic features and disease mechanisms. 
Applying our method can provide deeper insights into these complexities, ultimately leading to the development of more effective and personalized treatment plans.

\bibliographystyle{apalike}
\bibliography{bib1}

\newpage
\appendix

\centerline{\Large\bf Supplementary Materials}
\vspace{0.3in}

\setcounter{section}{0}
\setcounter{theorem}{0}
\setcounter{figure}{0}
\setcounter{table}{0}

\renewcommand{\thesection}{S\arabic{section}}
\renewcommand{\thetheorem}{S\arabic{theorem}}
\renewcommand{\thelemma}{S\arabic{lemma}}
\renewcommand{\thefigure}{S\arabic{figure}}
\renewcommand{\thetable}{S\arabic{table}}

\theoremstyle{plain}
\newtheorem{theoremS}{Theorem}[section]
\newtheorem{lemmaS}[theoremS]{Lemma}
\newtheorem{propositionS}[theoremS]{Proposition}

\theoremstyle{remark}
\newtheorem{remarkS}[theoremS]{Remark}


\section{Simulation settings}

\begin{table}[ht]
\centering 
\begin{tabular}{lrrr}
  \hline
 $\pi_{cj}$ & Class 1 ($c=1$) & Class 2 ($c=2$) & Class 3 ($c=3$) \\ 
  \hline
  Variable 1 ($j=1$) & 0.1 & 0.9 & 0.5 \\ 
  Variable 2 ($j=2$) & 0.5 & 0.1 & 0.9 \\ 
  Variable 3 ($j=3$) & 0.9 & 0.5 & 0.1 \\ 
  Variable 4 ($j=4$) & 0.1 & 0.9 & 0.5 \\ 
  Variable 5 ($j=5$) & 0.5 & 0.1 & 0.9 \\ 
   \hline
\end{tabular} 
\caption{\label{stab1} Prevalence values to generate binary variables. }
\end{table}

\begin{table}[ht]
\centering 
\begin{tabular}{lrrr}
  \hline
$\lambda_{kc}$ & Class 1 ($c=1$) & Class 2 ($c=2$) & Class 3 ($c=3$) \\
  \hline
Target ($k=0$)&0.50&0.30&0.20 \\
Site 1 ($k=1$)&0.45&0.35&0.20 \\
Site 2 ($k=2$)&0.55&0.25&0.20 \\
Site 3 ($k=3$)&0.45&0.20&0.35 \\
Site 4 ($k=4$)&0.55&0.20&0.25 \\
Site 5 ($k=5$)&0.50&0.20&0.30 \\
Site 6 ($k=1$)&0.45&0.35&0.20 \\
Site 7 ($k=2$)&0.55&0.25&0.20 \\
Site 8 ($k=3$)&0.45&0.20&0.35 \\
Site 9 ($k=4$)&0.55&0.20&0.25 \\
Site 10 ($k=5$)&0.50&0.20&0.30 \\

   \hline
\end{tabular} 
\caption{\label{stab2} Mixing proportions of the three subpopulations in the target study and source studies. Small differences exist in the mixing proportions of the target study and source studies. }
\end{table}

\begin{table}[ht]
\centering 
\begin{tabular}{lrrr}
  \hline
$\lambda_{kc}$ & Class 1 ($c=1$) & Class 2 ($c=2$) & Class 3 ($c=3$) \\
  \hline
Target ($k=0$)&0.80&0.10&0.10 \\
Site 1 ($k=1$)&0.10&0.10&0.80 \\
Site 2 ($k=2$)&0.11&0.09&0.80 \\
Site 3 ($k=3$)&0.09&0.11&0.80 \\
Site 4 ($k=4$)&0.10&0.11&0.79 \\
Site 5 ($k=5$)&0.11&0.10&0.79 \\
Site 6 ($k=6$)&0.12&0.10&0.78 \\
Site 7 ($k=7$)&0.10&0.12&0.78 \\
Site 8 ($k=8$)&0.09&0.10&0.81 \\
Site 9 ($k=9$)&0.10&0.09&0.81 \\
Site 10 ($k=10$)&0.09&0.09&0.82 \\
   \hline
\end{tabular} 
\caption{\label{stab3} Mixing proportions of the three subpopulations in the target study and source studies. Significant differences exist in the mixing proportions of the target study and source studies. }
\end{table}

\clearpage

\section{Additional simulation results}

\begin{figure}[!t]
    \centering
    \includegraphics[width=0.8\textwidth]{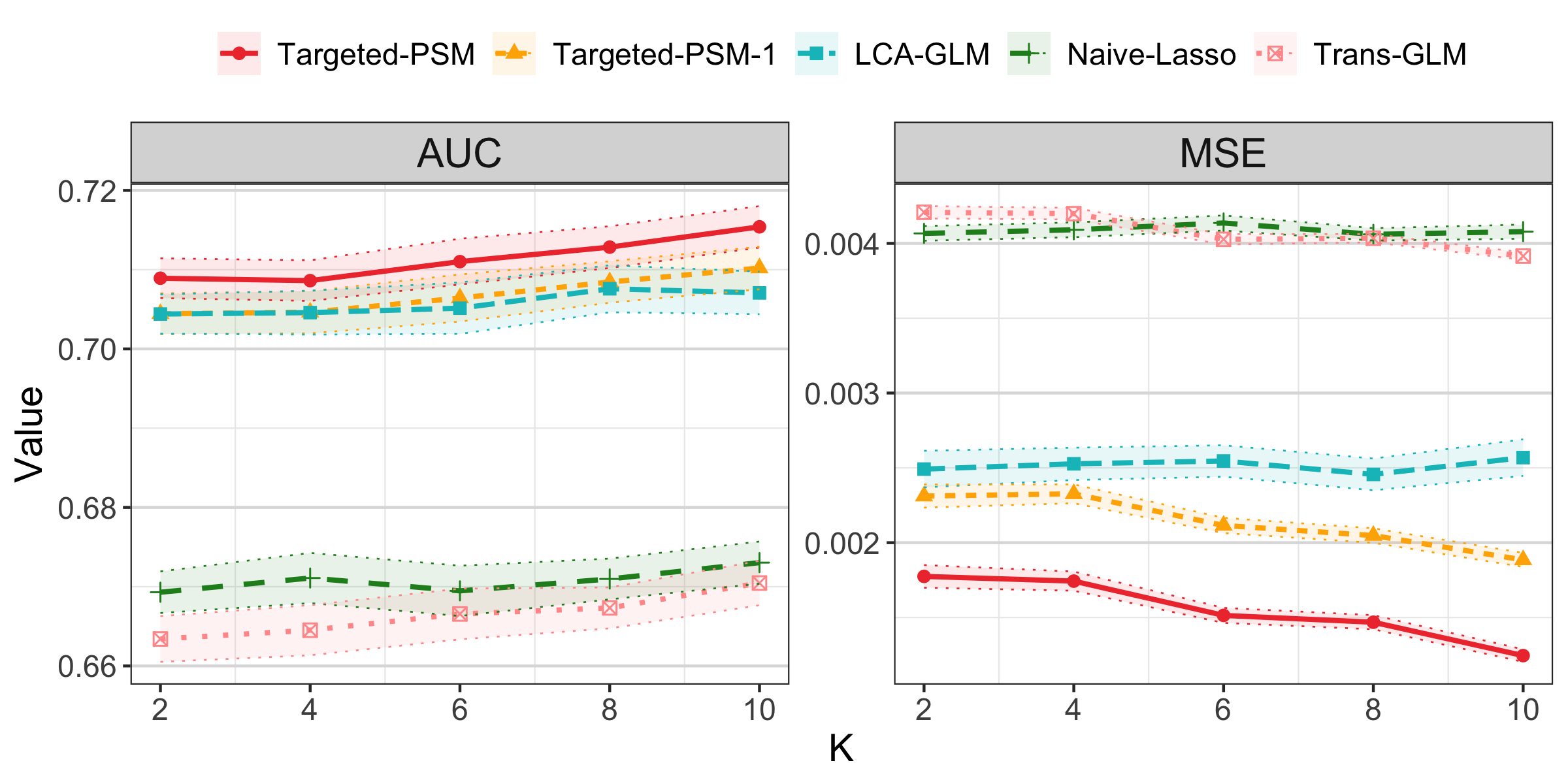}
    \caption{The estimation and prediction results for different methods when the subpopulations are well separated. 
    We let $h=15$ and the difference in mixing proportions between the target study and source studies be small in this setting.}
    \label{fig:3}
\end{figure}

\begin{figure}[!t]
    \centering
    \includegraphics[width=0.8\textwidth]{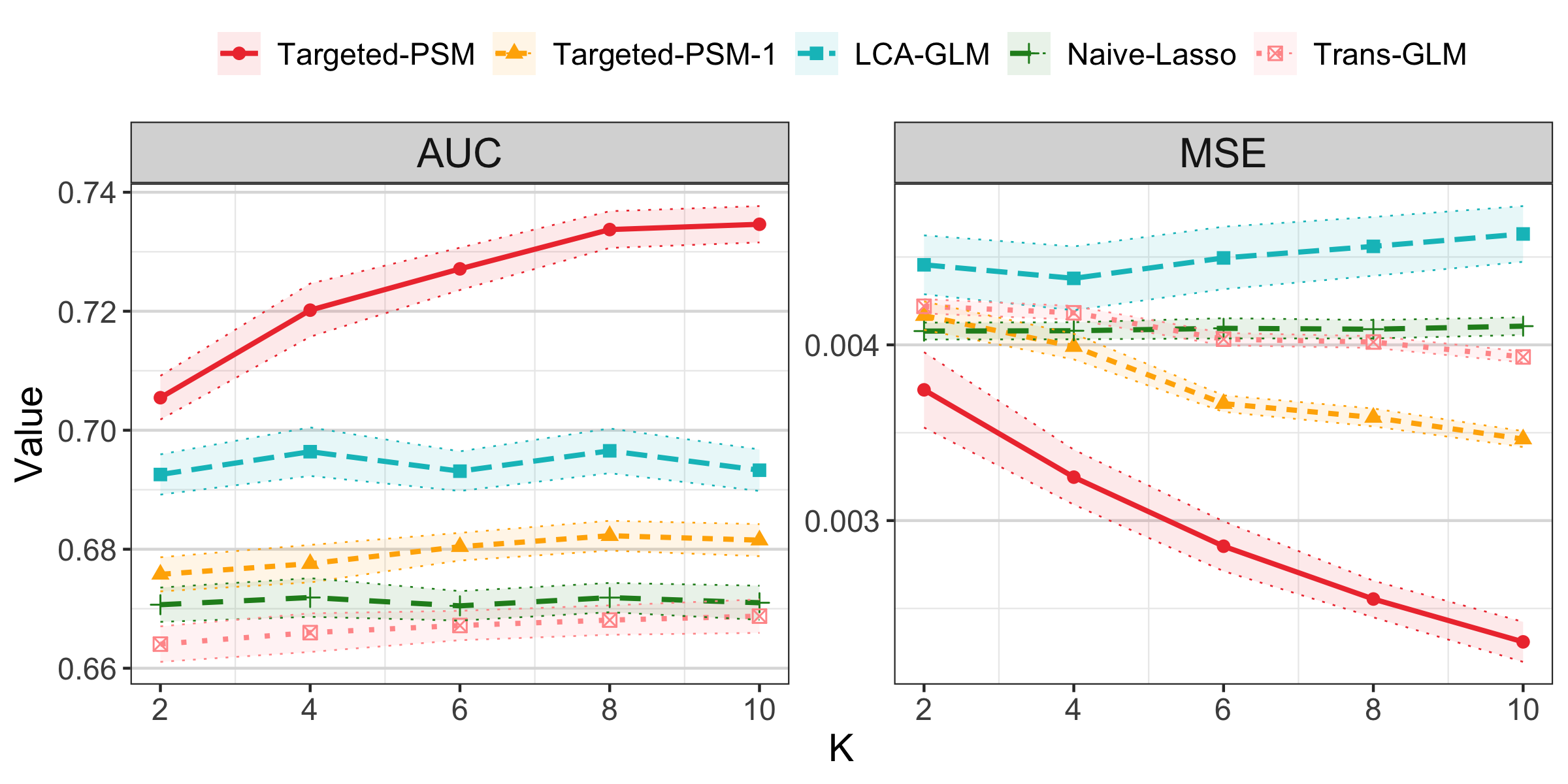}
    \caption{The estimation and prediction results for different methods when the subpopulations are not well separated. 
    We let $h=15$ and the difference in mixing proportions between the target study and source studies be small in this setting.}
    \label{fig:4}
\end{figure}

\begin{figure}[!ht]
    \centering
    \includegraphics[width=0.8\textwidth]{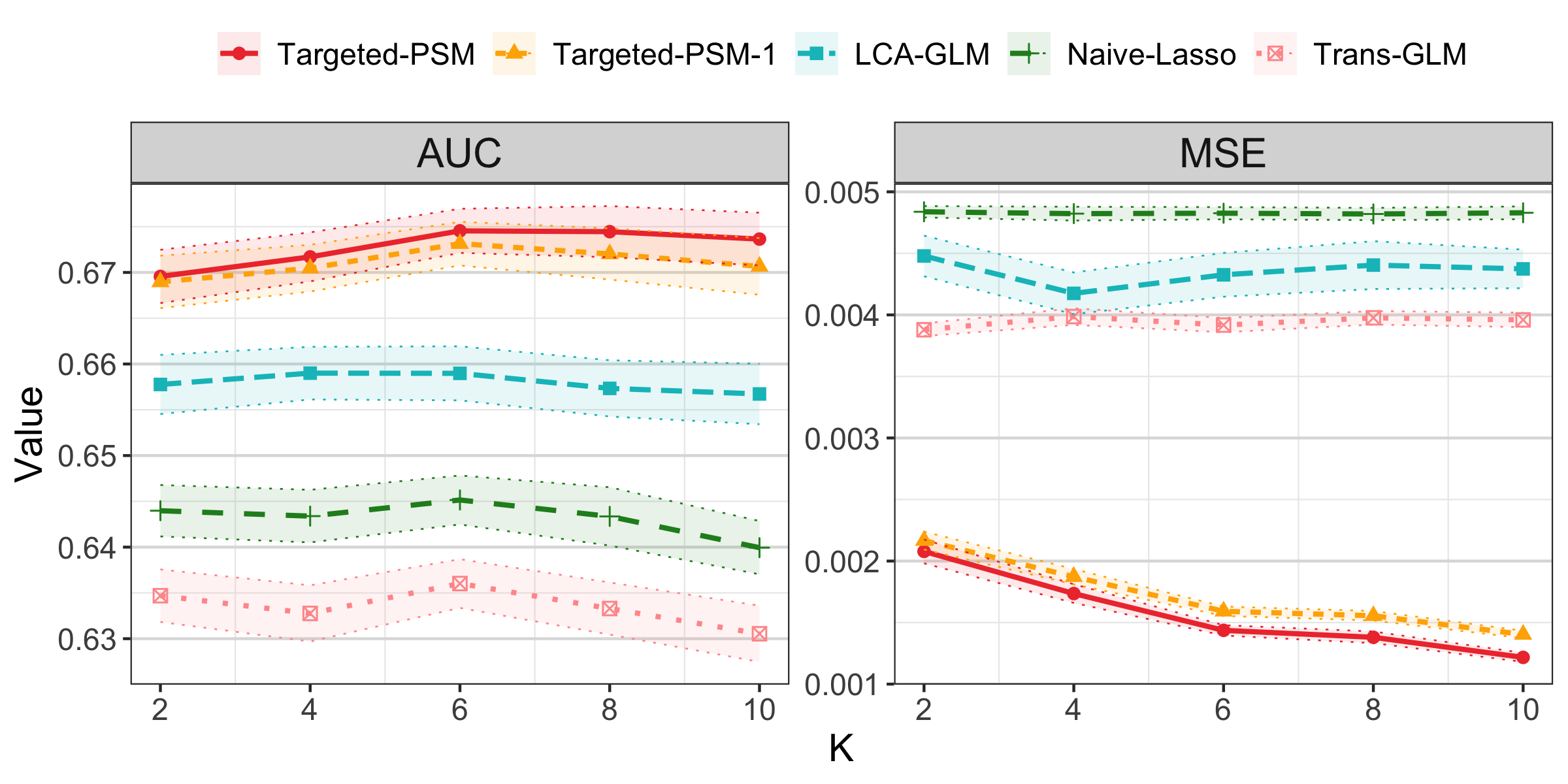}
    \caption{The estimation and prediction results for different methods when the subpopulations are well separated. 
    We let $h=15$ and the difference in mixing proportions between the target study and source studies be large in this setting.}
    \label{fig:7}
\end{figure}

\begin{figure}[!ht]
	\centering
	\includegraphics[width=0.8\textwidth]{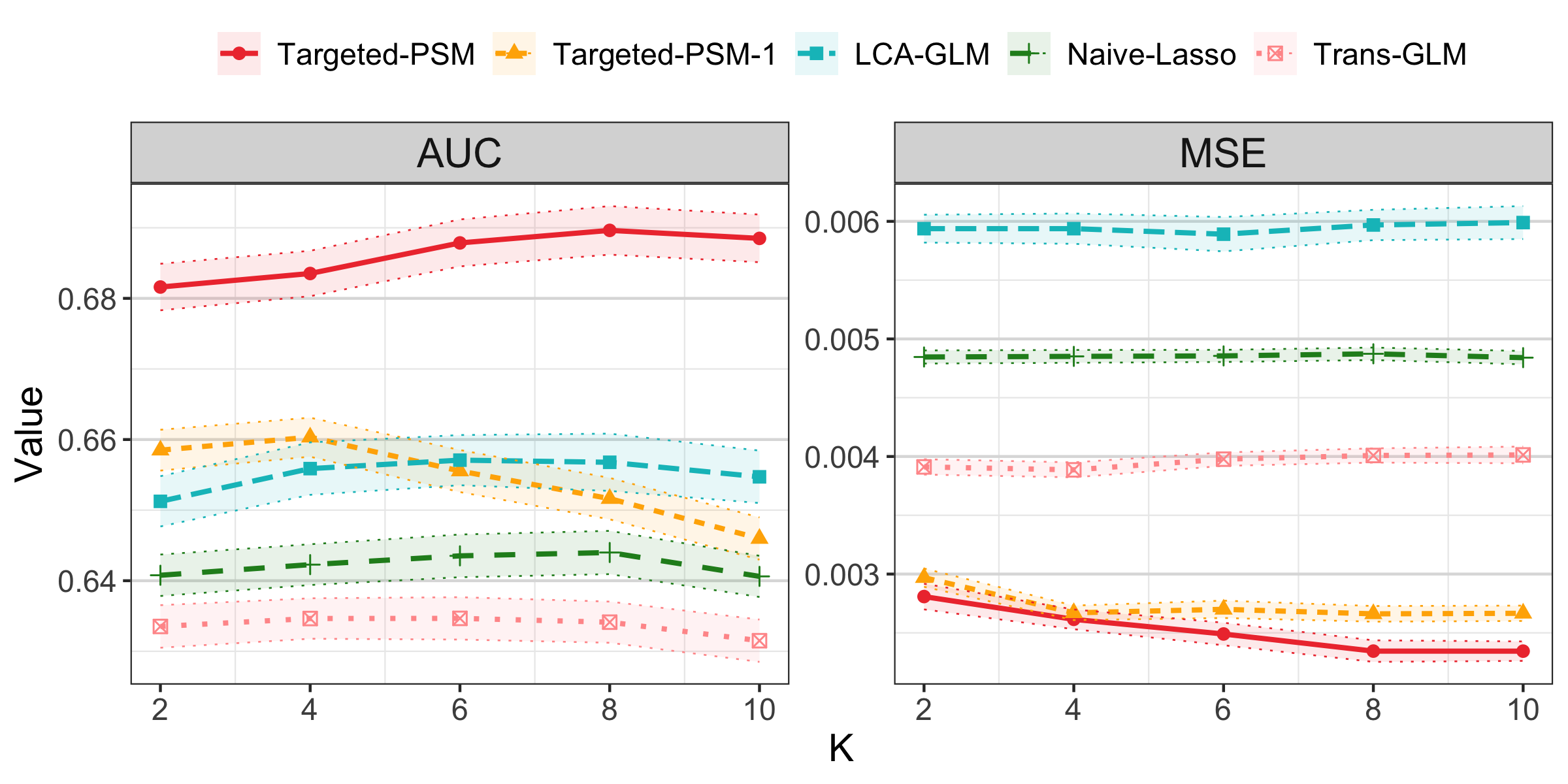}
	\caption{The estimation and prediction results for different methods when the subpopulations are not well separated. 
    We let $h=15$ and the difference in mixing proportions between the target study and source studies be large in this setting.}
	\label{fig:8}
\end{figure}




\end{document}